\RequirePackage{lineno}
\documentclass[prl,twocolumn,a4paper,superscriptaddress,showpacs,showkeys]{revtex4}
\usepackage{graphicx}
\usepackage{dcolumn}   % needed for some tables
\usepackage{amsmath}% loads amstext, amsbsy, amsopn but not amssymb
\usepackage{enumerate}
\usepackage{placeins}
\usepackage{appendix}
\usepackage[T1]{fontenc}
\usepackage[utf8]{inputenc}
\usepackage{chngcntr}
\usepackage{microtype}
\usepackage{lineno}
\usepackage{color}
\usepackage{xspace}
\usepackage{lipsum,mathptmx,etoolbox}
\usepackage{enumerate}

\usepackage{caption}

\usepackage{enumerate}
\usepackage{colortbl}
\definecolor{darkred}{rgb}{0.5,0,0}
\definecolor{darkblue}{rgb}{0,0,0.5}
\definecolor{firebrick}{rgb}{0.75,0.125,0.125}
\definecolor{darkgreen}{rgb}{0,0.5,0}
\definecolor{green}{rgb}{0,0.5,0}
\definecolor{red}{rgb}{0.5,0,0}

\newcommand{\eV}{\ensuremath{\mbox{e\kern-0.1em V}}\xspace}
\newcommand{\GeV}{\ensuremath{\mbox{Ge\kern-0.1em V}}\xspace}
\newcommand{\MeV}{\ensuremath{\mbox{Me\kern-0.1em V}}\xspace}
\newcommand{\GeVc}{\ensuremath{\mbox{Ge\kern-0.1em V}\!/\!c}\xspace}
\newcommand{\GeVcc}{\ensuremath{\mbox{Ge\kern-0.1em V}\!/\!c^2}\xspace}
\newcommand{\AGeV}{\ensuremath{A\,\mbox{Ge\kern-0.1em V}}\xspace}
\newcommand{\AGeVc}{\ensuremath{A\,\mbox{Ge\kern-0.1em V}\!/\!c}\xspace}
\newcommand{\MeVc}{\ensuremath{\mbox{Me\kern-0.1em V}/c}\xspace}

\newcommand{\pt}{\ensuremath{p_{\mathrm{T}}}\xspace}

\newcommand{\mt}{\ensuremath{m_{\mathrm{T}}}\xspace}

\newcommand{\pim}{\ensuremath{\pi^-}\xspace}
\newcommand{\pip}{\ensuremath{\pi^+}\xspace}
\newcommand{\km}{\ensuremath{\textup{K}^-}\xspace}
\newcommand{\kp}{\ensuremath{\textup{K}^+}\xspace}

\newcommand{\CernVM}{\textsc{Cern\-\kern-0.05emVM}\xspace}
\newcommand{\NASixtyOne}{NA61\slash SHINE\xspace}%this seems to work properly to me. aa
\begin{document}
%\linenumbers

%\preprint{}

\title{Proton-proton interactions and onset of deconfinement}


%\author{}
%\thanks{}
%\affiliation{}


\date{\today}

\affiliation{National Nuclear Research Center, Baku, Azerbaijan}
\affiliation{Faculty of Physics, University of Sofia, Sofia, Bulgaria}
\affiliation{Ru{\dj}er Bo\v{s}kovi\'c Institute, Zagreb, Croatia}
\affiliation{LPNHE, University of Paris VI and VII, Paris, France}
\affiliation{Karlsruhe Institute of Technology, Karlsruhe, Germany}
\affiliation{University of Frankfurt, Frankfurt, Germany}
\affiliation{Wigner Research Centre for Physics of the Hungarian Academy of Sciences, Budapest, Hungary}
\affiliation{University of Bergen, Bergen, Norway}
\affiliation{Jan Kochanowski University in Kielce, Poland}
\affiliation{Institute of Nuclear Physics, Polish Academy of Sciences, Cracow, Poland}
\affiliation{National Centre for Nuclear Research, Warsaw, Poland}
\affiliation{Jagiellonian University, Cracow, Poland}
\affiliation{AGH - University of Science and Technology, Cracow, Poland}
\affiliation{University of Silesia, Katowice, Poland}
\affiliation{University of Warsaw, Warsaw, Poland}
\affiliation{University of Wroc{\l}aw,  Wroc{\l}aw, Poland}
\affiliation{Warsaw University of Technology, Warsaw, Poland}
\affiliation{Institute for Nuclear Research, Moscow, Russia}
\affiliation{Joint Institute for Nuclear Research, Dubna, Russia}
\affiliation{National Research Nuclear University (Moscow Engineering Physics Institute), Moscow, Russia}
\affiliation{St. Petersburg State University, St. Petersburg, Russia}
\affiliation{University of Belgrade, Belgrade, Serbia}
\affiliation{University of Geneva, Geneva, Switzerland}
\affiliation{Fermilab, Batavia, USA}
\affiliation{University of Colorado, Boulder, USA}
\affiliation{University of Pittsburgh, Pittsburgh, USA}

\author{A.~\surname{Aduszkiewicz}}
\affiliation{University of Warsaw, Warsaw, Poland}
\author{E.V.~\surname{Andronov}}
\affiliation{St. Petersburg State University, St. Petersburg, Russia}
\author{T.~\surname{Anti\'ci\'c}}
\affiliation{Ru{\dj}er Bo\v{s}kovi\'c Institute, Zagreb, Croatia}
\author{V.~\surname{Babkin}}
\affiliation{Joint Institute for Nuclear Research, Dubna, Russia}
\author{M.~\surname{Baszczyk}}
\affiliation{AGH - University of Science and Technology, Cracow, Poland}
\author{S.~\surname{Bhosale}}
\affiliation{Institute of Nuclear Physics, Polish Academy of Sciences, Cracow, Poland}
\author{A.~\surname{Blondel}}
\affiliation{LPNHE, University of Paris VI and VII, Paris, France}
\author{M.~\surname{Bogomilov}}
\affiliation{Faculty of Physics, University of Sofia, Sofia, Bulgaria}
\author{A.~\surname{Brandin}}
\affiliation{National Research Nuclear University (Moscow Engineering Physics Institute), Moscow, Russia}
\author{A.~\surname{Bravar}}
\affiliation{University of Geneva, Geneva, Switzerland}
% ...

\collaboration{\NASixtyOne Collaboration}
\noaffiliation

\begin{abstract}

The \NASixtyOne experiment at the CERN SPS is performing a uniqe study
of the phase diagram of strongly interacting matter by varying
collision energy and nuclear mass number of colliding nuclei. In
central Pb+Pb collisions the NA49 experiment found structures in
the energy dependence of several observables in the CERN SPS energy
range that had been predicted for the transition to a deconfined phase.
New measurements of \NASixtyOne find intriguing similarities
in p+p interactions for which no deconfinement transition is expected
at SPS energies. Possible implications will be discussed.

\end{abstract}


%\keywords{}

\maketitle


%\subsection{}
%\subsubsection{}

%
%
%%***********************************************************************************
%\section{Introduction}

The standard approach to heavy-ion collisions~\cite{Florkowski:2010zz} assumes creation of 
strongly interacting matter in local equilibrium at the early stage of a collision. 
The matter properties depend on energy and baryon densities via an equation-of-state. The matter 
expansion is modelled by hydrodynamics and its conversion to final state hadrons by 
statistical hadronization models~\cite{Torrieri:2004zz, Becattini:2005xt, Andronic:2017pug}.
The early stage energy density monotonically increases with collision energy and at 
sufficiently high energies the state of matter is expected to change from the confined phase to 
the quark-gluon plasma (QGP). 

In an energy scan of central Pb+Pb collisions at the CERN SPS the NA49 experiment 
found structures in a common narrow energy 
interval $\sqrt{s_{NN}} \approx$ 7-12~GeV ~\cite{Afanasiev:2002mx,Alt:2007aa,Gazdzicki:2010iv} 
($\sqrt{s_{NN}}$ is the collision energy per nucleon pair in the centre of mass system)
for several observables that had been predicted~\cite{Gazdzicki:1998vd} 
for the transition to the QGP phase. The most conclusive are:
\begin{enumerate}[(i)]
\item a fast rise and and sharp peak in the ratio of strangeness to entropy production
\item a fast rise and following plateau of the temperature as measured by the inverse slope
      parameter of the kaon transverse mass distributions
\end{enumerate}

Experimental results on p+p interactions served as an important reference with respect to which
new physics in heavy-ion collisions was searched for. 
The most popular models of p+p interactions are qualitatively different from the standard approach to 
heavy-ion collisions. They are resonance-string models~\cite{Andersson:1983ia} in which the hydrodynamic
expansion of the strongly interacting matter created in nucleus-nucleus (A+A) collisions is 
replaced in p+p collisions by excitation of resonances or strong fields between colour charges of 
quarks and di-quarks (strings). The assumption of statistical hadronization of matter 
is substituted by dynamical modelling of resonance and/or string decays as well as 
quark/gluon fragmentation into hadrons. Since the early days, the different modelling 
of p+p interactions and heavy-ion collisions was supported by qualitative
disagreement of the p+p data with predictions of statistical and hydrodynamical models - large 
particle multiplicity fluctuations and a power-law shape of transverse momentum spectra 
at high \pt~\cite{Begun:2008fm}. 
On the other hand, the different modelling has been questioned 
by striking agreement of the p+p data with other predictions of statistical and hydrodynamical 
models - mean multiplicities of hadrons and transverse mass spectra 
at low and intermediate \pt  follow a similar pattern.
%predicted by the models~\cite{Hagedorn:1965st,Becattini:2009sc}.
%}
Moreover, recent LHC data on the azimuthal angle distribution of charged particles in high multiplicity 
p+p interactions~\cite{CMS:2012qk,Khachatryan:2016txc,Nagle:2018nvi} show anisotropies up to now observed
only in heavy-ion collisions and attributed to the hydrodynamical expansion of matter~\cite{Bozek:2016acp}. 
Also it was reported that relative strange particle yields in p+p interactions at LHC smoothly increase 
with increasing charged particle multiplicity and for high multiplicity collisions are close to
those in Pb+Pb collisions~\cite{ALICE:2017jyt}.

%the Pb+Pb ones~\cite{ALICE:2017jyt}.
These results suggests that the observation of strongly interacting matter in nucleus-nucleus collisions
may also extend to those p+p interactions at LHC energies which produce sufficiently high particle multiplicity.

This paper addresses the relation between the observation of effects possibly indicating the onset of deconfinement in Pb+Pb collisions
and recently uncovered, still unexplained features in p+p interactions.

%\section{Results}
New experimental insight is possible thanks to recent results on p+p interactions 
at the CERN SPS from the \NASixtyOne~\cite{Abgrall:2014fa} fixed target large acceptance hadron detector. 
%The \NASixtyOne tracking system consists of 4 large volume time projection chambers (TPCs). 
%Two of the TPCs (VTPC1 and VTPC2) are
%placed inside superconducting dipole magnets. Downstream of the magnets two larger TPCs (MTPC-R and
%MTPC-L) provide acceptance at high momenta. The interactions were produced with a beam of
%158~\GeVc protons on a cylindrical liquid hydrogen target of 20~cm length and 2~cm transverse diameter.
The measurements~\cite{Aduszkiewicz:2017sei,Aduszkiewicz:2015jna} cover the energy range in which 
experimental effects attributed to the onset of deconfinement in heavy-ion collisions are located.
They allow to significantly extend and improve the world data on the 
\kp/\pip ratio~~\cite{Gazdzicki:1995zs,Gazdzicki:1996pk} and the inverse slope parameter $T$ 
of transverse mass spectra of kaons~\cite{Kliemant:2003sa}.
Furthermore, recent data on p+p interactions at LHC energies allow to establish 
the collision energy dependence of bulk hadron production properties in the energy range in which 
the quark-gluon plasma is likely to be created in heavy-ion collisions.

\begin{figure*}
%\fbox{
%\begin{minipage}{16 cm}
	\begin{center}
		\includegraphics[width=0.48\textwidth]{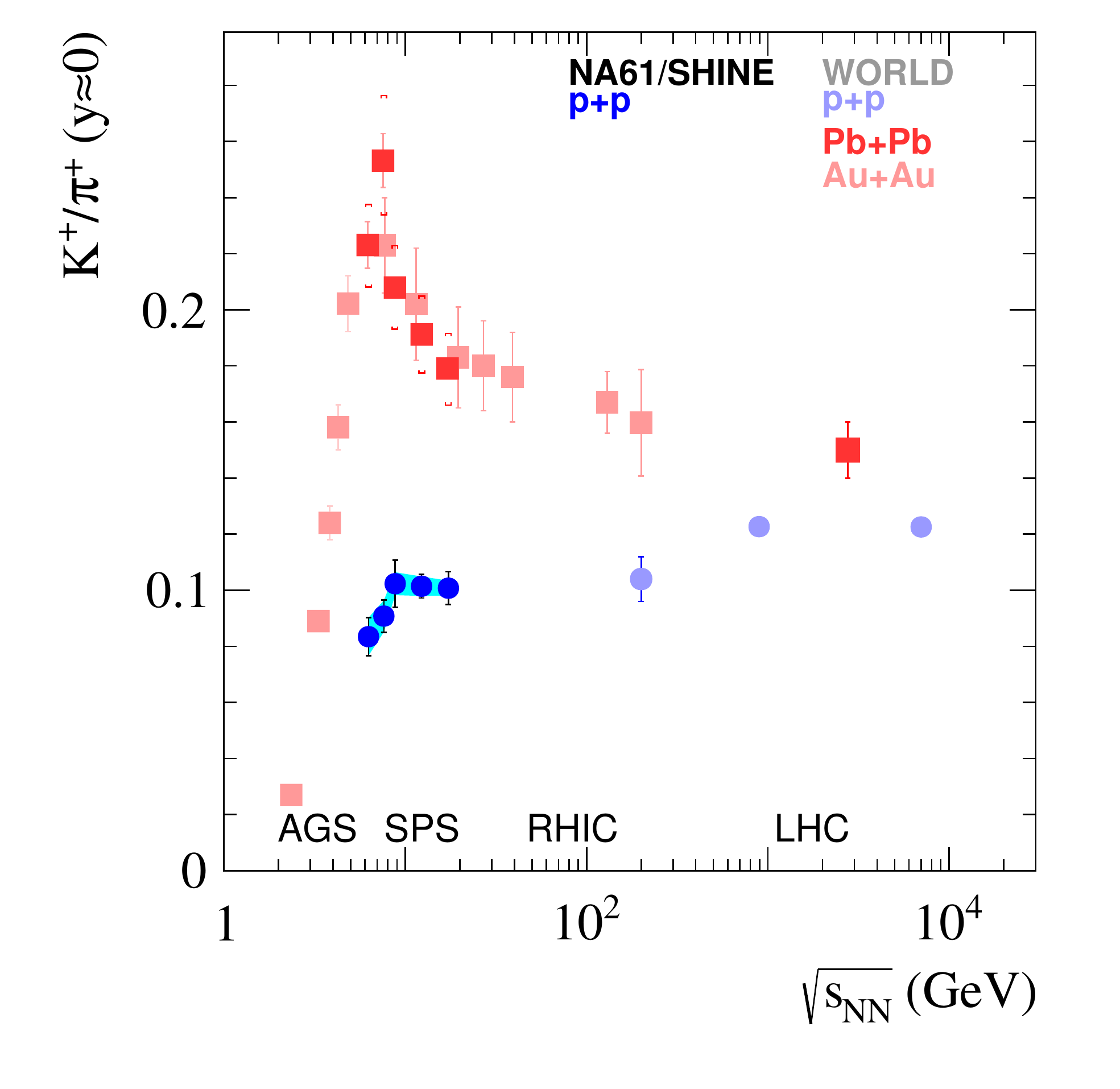}
		\includegraphics[width=0.48\textwidth]{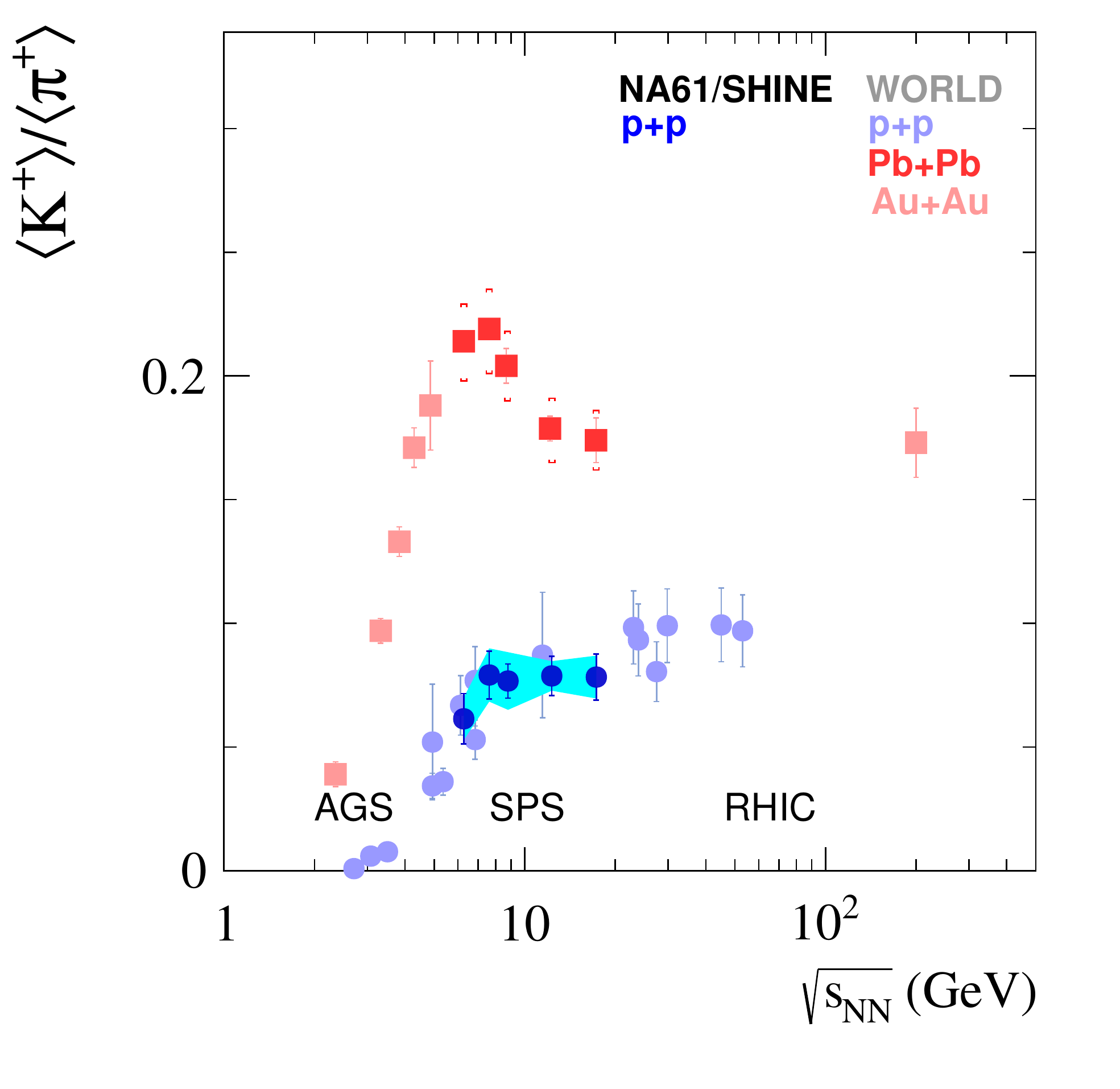}	\\
		\includegraphics[width=0.48\textwidth]{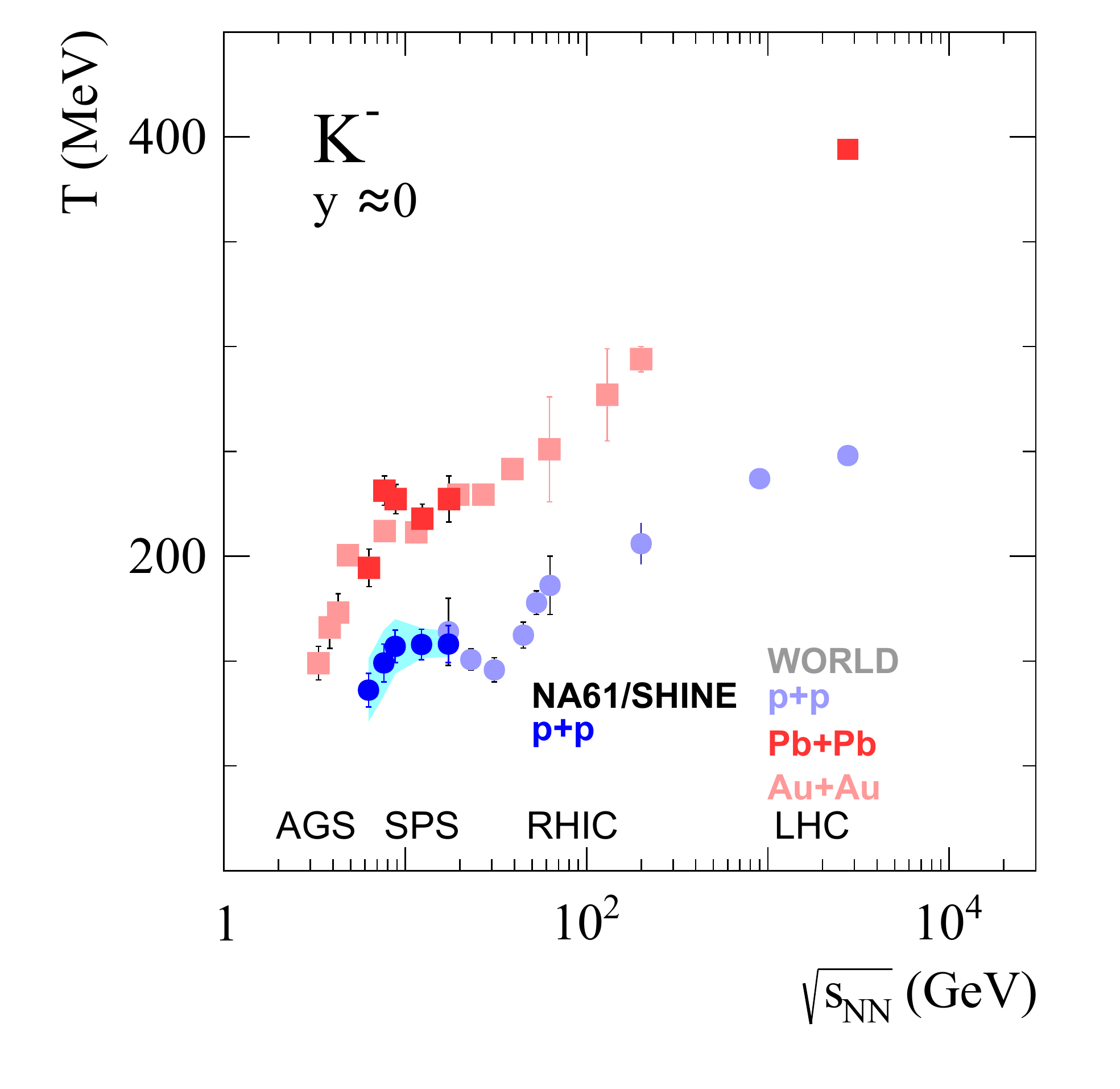}
		\includegraphics[width=0.48\textwidth]{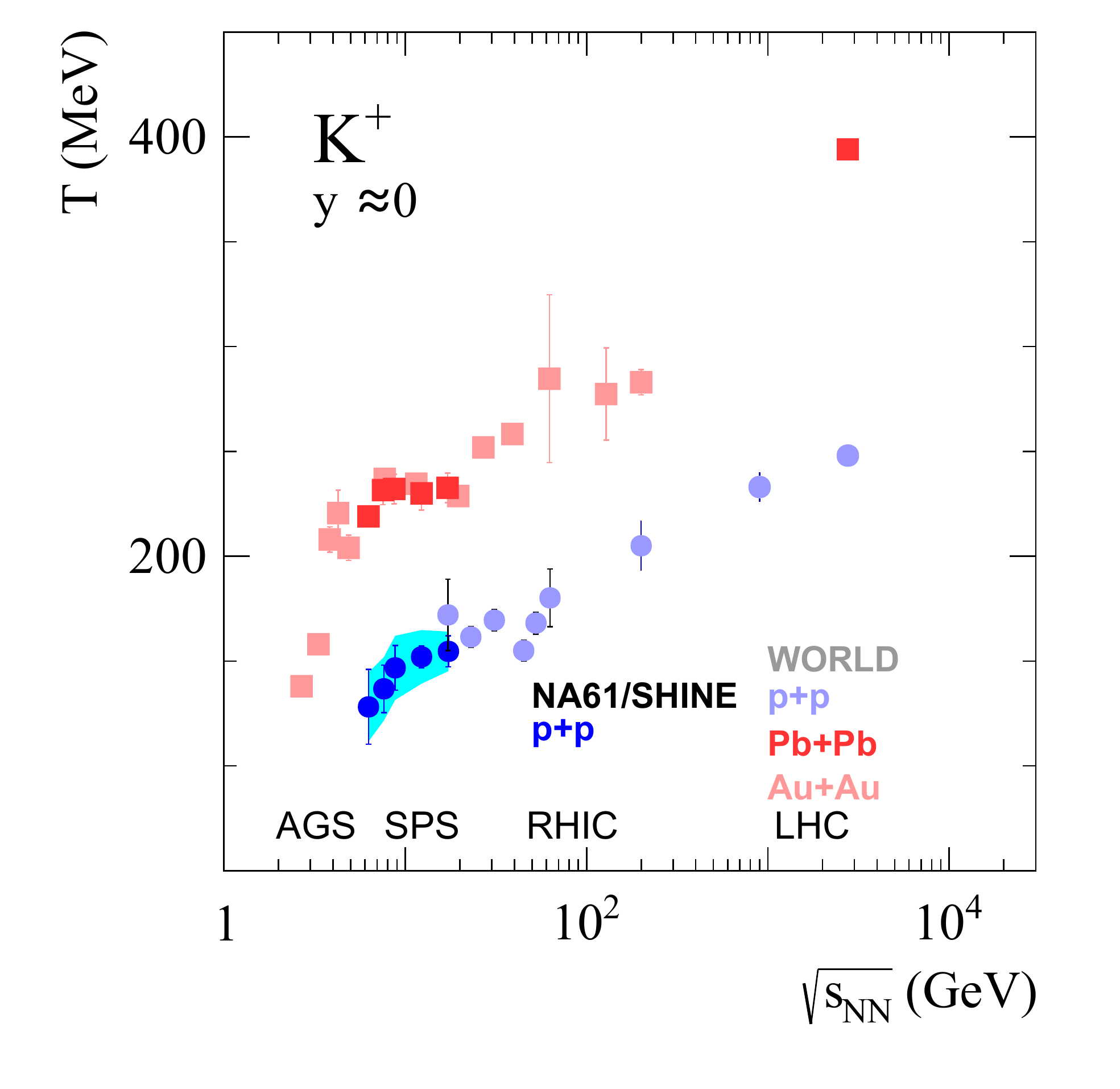}
	\end{center}
	\caption{Energy dependence of the \kp/\pip ratio at mid-rapidity (\emph{top-left}) and in the full phase-space (\emph{top-right}) as well as the inverse slope parameter $T$ of transverse mass spectra at mid-rapidity for \km (\emph{bottom-left}) and \kp  (\emph{bottom-right}) mesons. 
	The \NASixtyOne results for inelastic p+p interactions are shown together with the world data on p+p interactions as well as central Pb+Pb and Au+Au collisions~{\protect{\cite{Kliemant:2003sa,Aamodt:2011zj,Alper:1975jm,Abelev:2008ab,Abelev:2014laa,Ahle:1998jc,Ahle:1998gv,Ahle:1999va,Ahle:1999uy,Ahle:2000wq,Barrette:1999ry,Pelte:1997rg,Bearden:2004yx,Adamczyk:2017iwn,Abelev:2013vea,Adam:2015qaa}}}. Shaded bands show the systematic uncertainty.}
	\label{fig:ppPbPb}
%	\end{minipage}
%}
\end{figure*}

The energy dependence of the \kp/\pip ratio at mid-rapidity and in the full phase-space for 
inelastic p+p interactions is shown in Fig.~\ref{fig:ppPbPb} \emph{top-left} and \emph{top-right}, 
respectively. The results for heavy-ion (Pb+Pb and Au+Au) collisions are plotted for comparison.
The \NASixtyOne results on p+p interactions at CERN SPS energies are shown together with 
the world data~\cite{Gazdzicki:1995zs,Gazdzicki:1996pk,Arsene:2005mr,Aamodt:2011zj,Abelev:2012wca,Alper:1975jm,Abelev:2008ab,Abelev:2014laa,Ahle:1998jc,Ahle:1998gv,Ahle:1999va,Ahle:1999uy,Ahle:2000wq,Barrette:1999ry,Pelte:1997rg,Bearden:2004yx,Adamczyk:2017iwn,Abelev:2013vea,Adam:2015qaa}.
Results on the mid-rapidity ratio (the top-left plot) cover the range from low SPS energy to LHC energies.
For comparison the mid-rapidity plot includes also p+p data of other experiments on the full phase-space ratio.
The p+p data on the full phase-space ratio (the top-right plot) extends only to $\sqrt{s_{NN}} \approx 50$~GeV, 
whereas the heavy-ion data reach 200~GeV. 
The energy dependence of the mid-rapidity and full phase-space ratio in inelastic p+p interactions 
is similar. This seems to be also true for heavy-ion collisions. 

%%% Horn
The collision energy dependence of the \kp/\pip ratio in heavy-ion collisions shows the so-called \textit{horn} structure.
Following a fast rise the ratio passes through a maximum in
the SPS range and then settles to a plateau value at higher energies.

\begin{figure*}
%\fbox{
%\begin{minipage}{16 cm}

	\begin{center}
		\includegraphics[width=0.48\textwidth]{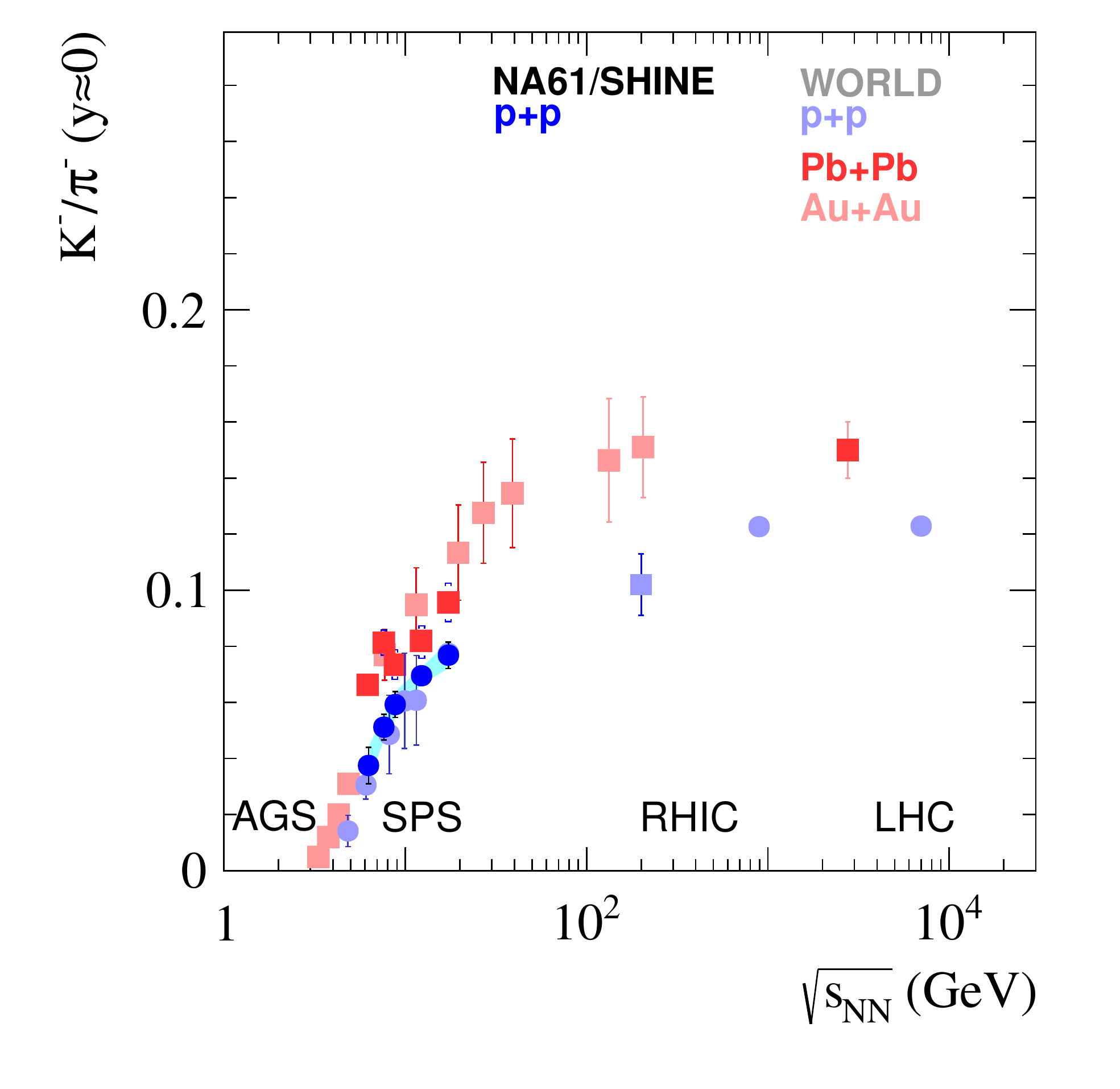}
		\includegraphics[width=0.48\textwidth]{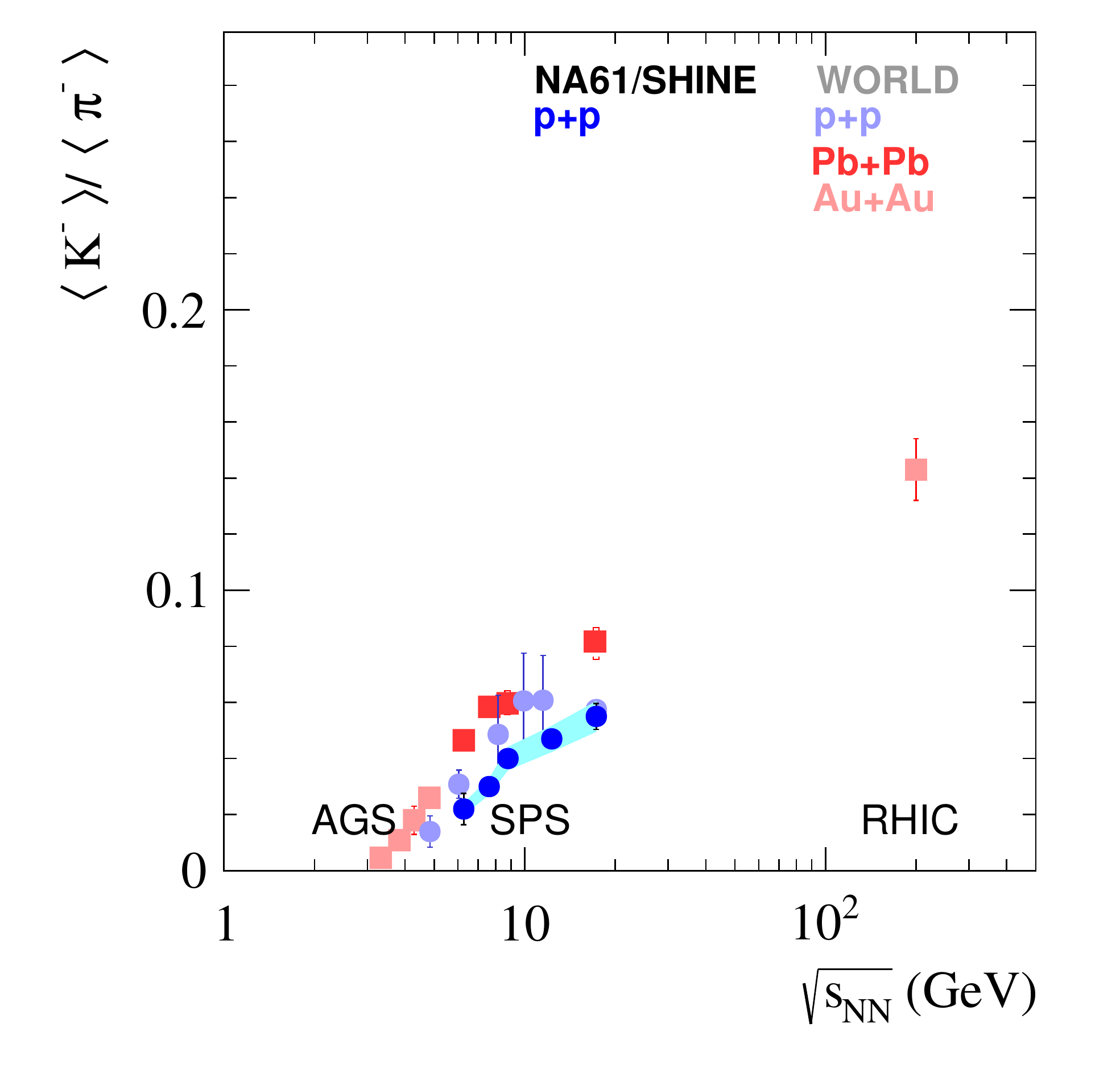}	\\
	\end{center}
\caption{
		Energy dependence of the \km/\pim ratio at mid-rapidity (\emph{left}) and in the full phase-space (\emph{right}). 
		The \NASixtyOne results for inelastic p+p interactions are shown together with the world data on p+p interactions as well as central Pb+Pb and Au+Au collisions~{\protect{\cite{Kliemant:2003sa,Aamodt:2011zj,Alper:1975jm,Abelev:2008ab,Abelev:2014laa,Ahle:1998jc,Ahle:1998gv,Ahle:1999va,Ahle:1999uy,Ahle:2000wq,Barrette:1999ry,Pelte:1997rg,Bearden:2004yx,Adamczyk:2017iwn,Abelev:2013vea,Adam:2015qaa}}}. Shaded bands show
the systematic uncertainty.}
	\label{fig:ppPbPb-}
%	\end{minipage}
%}
\end{figure*}

The \kp/\pip ratio at SPS energies was shown to be a good measure of the strangeness to entropy 
ratio~\cite{Gazdzicki:2010iv} which is different in the confined phase (hadrons) and the QGP (quarks, anti-quarks and gluons). 
This is because at high baryon to meson ratio (SPS energies and below) the anti-hyperon yield is small and the main carriers of anti-strange quarks are \kp and $\mathrm{K}^0$ with $\langle \kp \rangle \approx \langle \mathrm{K}^0 \rangle$ due to
approximate isospin symmetry in heavy ion collisions. Thus the \kp yield counts about half of the strange 
quark - anti-quark pairs ($\langle s\bar{s} \rangle$) produced in 
the collisions and contained in the reaction products~\cite{Gazdzicki:2010iv}.
In contrast, fractions of strange quarks carried by \km, $\bar{\textrm K^0}$ and hyperons
are comparable and change significantly with the baryon to meson ratio.
At lower collision energies  relatively more 
strange quarks are carried by hyperons and less by anti-kaons.  
Thus the energy dependence of the \km yield does not follow the energy dependence of $\langle s\bar{s} \rangle$.
This is illustrated in Fig.~\ref{fig:ppPbPb-} where the energy dependence of the 
\km/\pim ratio at mid-rapidity (the left plot) and in the full phase-space (the right plot) for inelastic p+p interactions
and heavy-ion collisions is shown.
In conclusion, the \kp yield is preferred over \km and $\Lambda$ yields when the total number of $s\bar{s}$ pairs is
of interest as in the search for the QGP~\cite{Koch:1986ud} and the onset of deconfinement~\cite{Gazdzicki:1998vd}.

Further comment is in order here.
Since many years
it has been popular to fit mean hadron multiplicities, which include multiplicities of kaons and pions, 
assuming that a hadron gas in equilibrium is created when strongly interacting matter hadronizes. 
The temperature, the baryon chemical potential, and the
hadronization volume are free parameters of the model and are fitted to the data at each
energy. In this formulation, the hadron gas model cannot make any prediction about the
energy dependence of hadron production so that an extension of the model was proposed,
in which the values of the temperature and baryon chemical potential evolve smoothly
with collision energy (see Ref.~\cite{Andronic:2017pug} for a recent review).
By construction (fits to the energy dependence of data on mean hadron multiplicities), the prevailing trend in the data is reproduced by the models.
This parametrization of the measured energy dependence of hadron yields
is often confused with predicting the energy dependence without invoking the phase transition.

The collision energy dependence of the \kp/\pip ratio in inelastic p+p interactions is different from the
one in heavy-ion collisions, see Fig.~\ref{fig:ppPbPb}. First of all, the ratio is smaller in p+p interactions than in
Pb+Pb and Au+Au collisions and does not show the horn structure. 

The p+p ratio approaches that in heavy-ion reactions with increasing energy, 
at LHC it is only about 10\% smaller than the corresponding ratio for central Pb+Pb collisions.
Starting from the threshold energy the ratio in p+p interactions steeply increases to reach a plateau at CERN SPS energies.
The plateau is followed by a weak increase towards LHC energies.
Notably, the beginning of the plateau in p+p interactions coincides with the horn maximum in
heavy-ion collisions. 

%%% Step
According to the standard model of heavy ion collisions the inverse slope parameter $T$ 
obtained from exponential fits of transverse mass spectra
is sensitive to both the temperature and the radial flow in the final state. 
The energy dependence of $T$ of transverse mass spectra of 
\kp  and  \km mesons produced at mid-rapidity in inelastic p+p interactions
is presented in Fig.~\ref{fig:ppPbPb}  \emph{bottom-left} and \emph{bottom-right}, respectively.
The \NASixtyOne results~\cite{Aduszkiewicz:2017sei} are
compared to the world data for p+p and heavy-ion
collisions~\cite{Kliemant:2003sa,Aamodt:2011zj,Alper:1975jm,Abelev:2008ab,Abelev:2014laa,Ahle:1998jc,Ahle:1998gv,Ahle:1999va,Ahle:1999uy,Ahle:2000wq,Barrette:1999ry,Pelte:1997rg,Bearden:2004yx,Adamczyk:2017iwn,Abelev:2013vea,Adam:2015qaa}.
Unless the T parameter was given directly by the experiment, it was taken from Ref.~\cite{Kliemant:2003sa}
or determined from transverse mass/momentum spectra according to the procedure of Ref.~\cite{Kliemant:2003sa}. The collision energy dependence of the $T$ parameter in heavy-ion collisions 
shows the so-called \textit{step} structure.
Following a fast rise the $T$ parameter passes through a stationary region (or even a weak minimum for \km), 
which starts at the low SPS energies, and then (above the top SPS energy)  
enters a domain of a steady increase. The increase continues up to the top LHC energy.
The step was predicted as a signal of the onset of deconfinement~\cite{Gazdzicki:1998vd,Gorenstein:2003cu}
resulting from the softening of the equation of state in the transition region.
The collision energy dependence of the $T$ parameter in inelastic p+p interactions 
is similar to the one for central Pb+Pb and Au+Au collisions.
The main difference is that the $T$ parameter in p+p interactions is significantly smaller than
for heavy-ion collisions which is usually attributed to smaller radial flow.

\begin{figure*}
%\fbox{
%\begin{minipage}{16 cm}

	\begin{center}
		\includegraphics[width=0.48\textwidth]{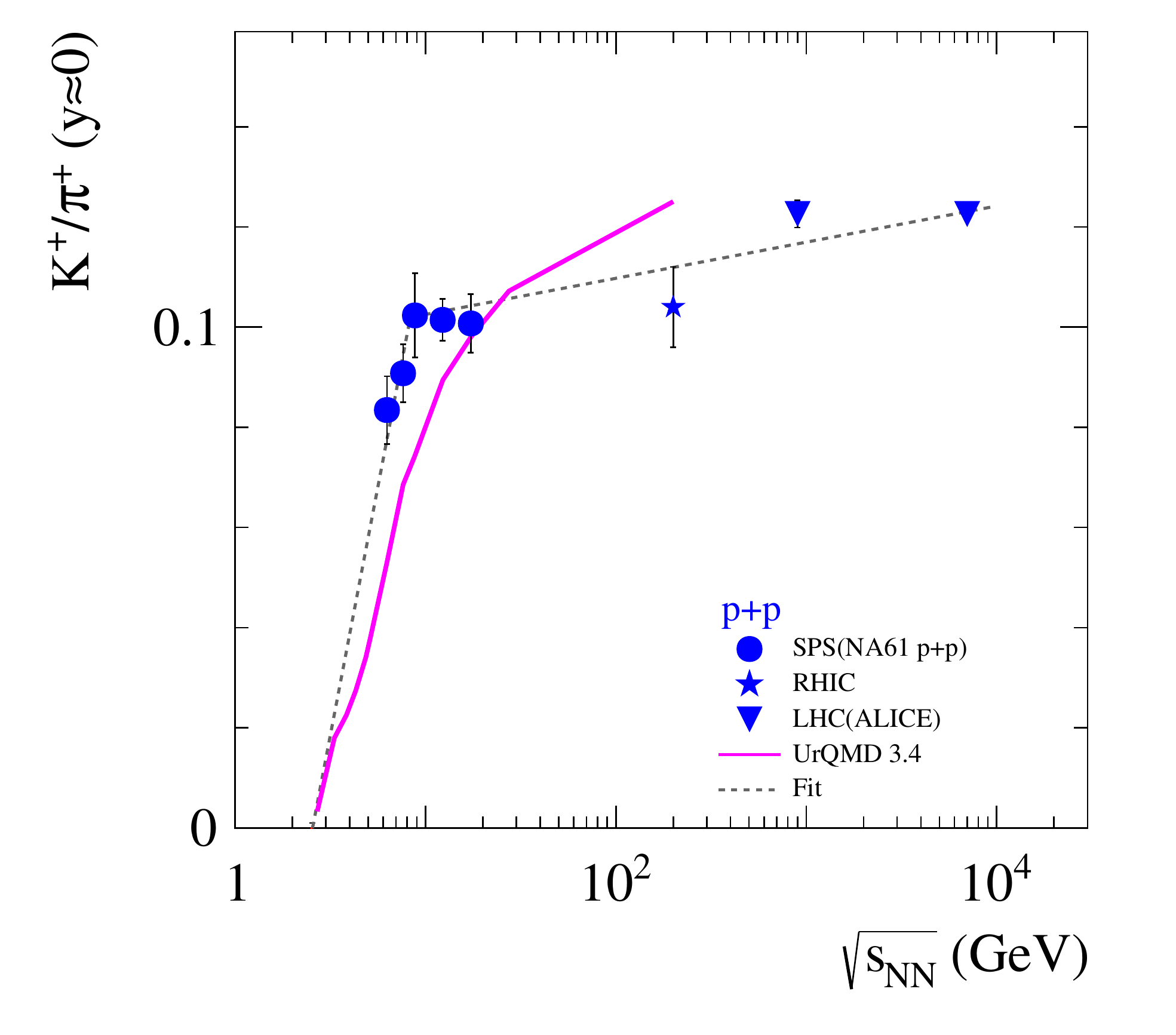}
		\includegraphics[width=0.48\textwidth]{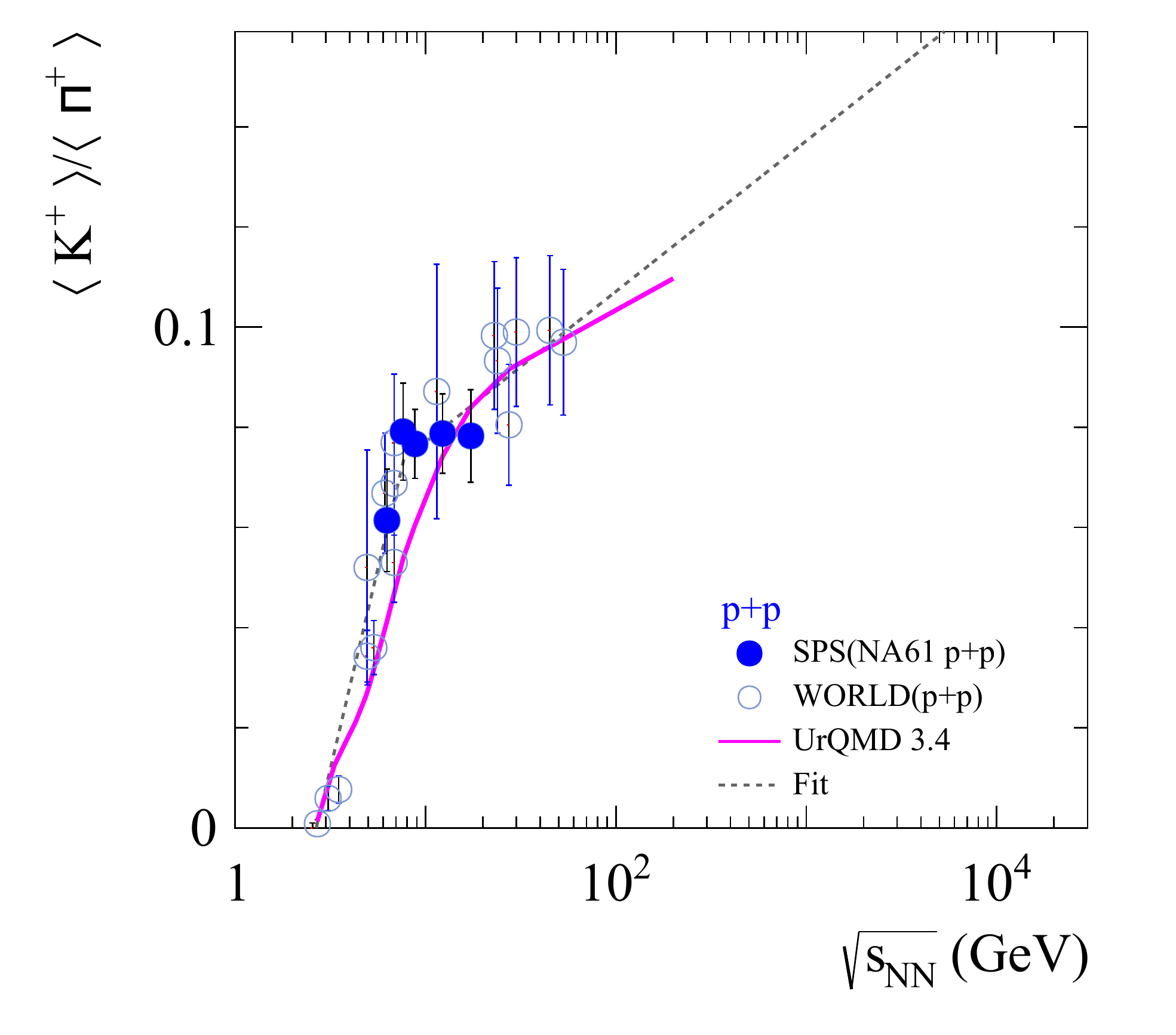}\\
		\includegraphics[width=0.48\textwidth]{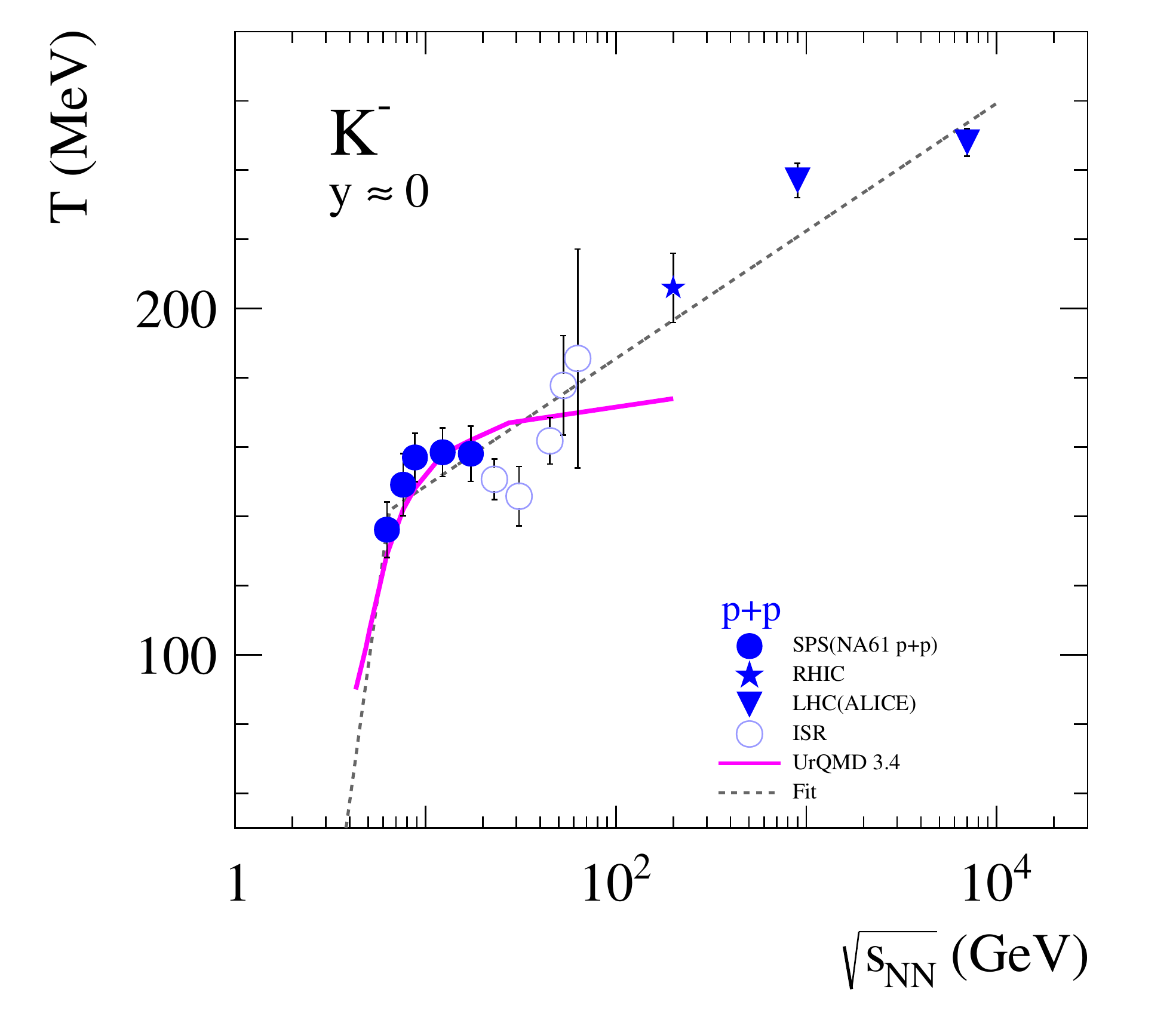}
		\includegraphics[width=0.48\textwidth]{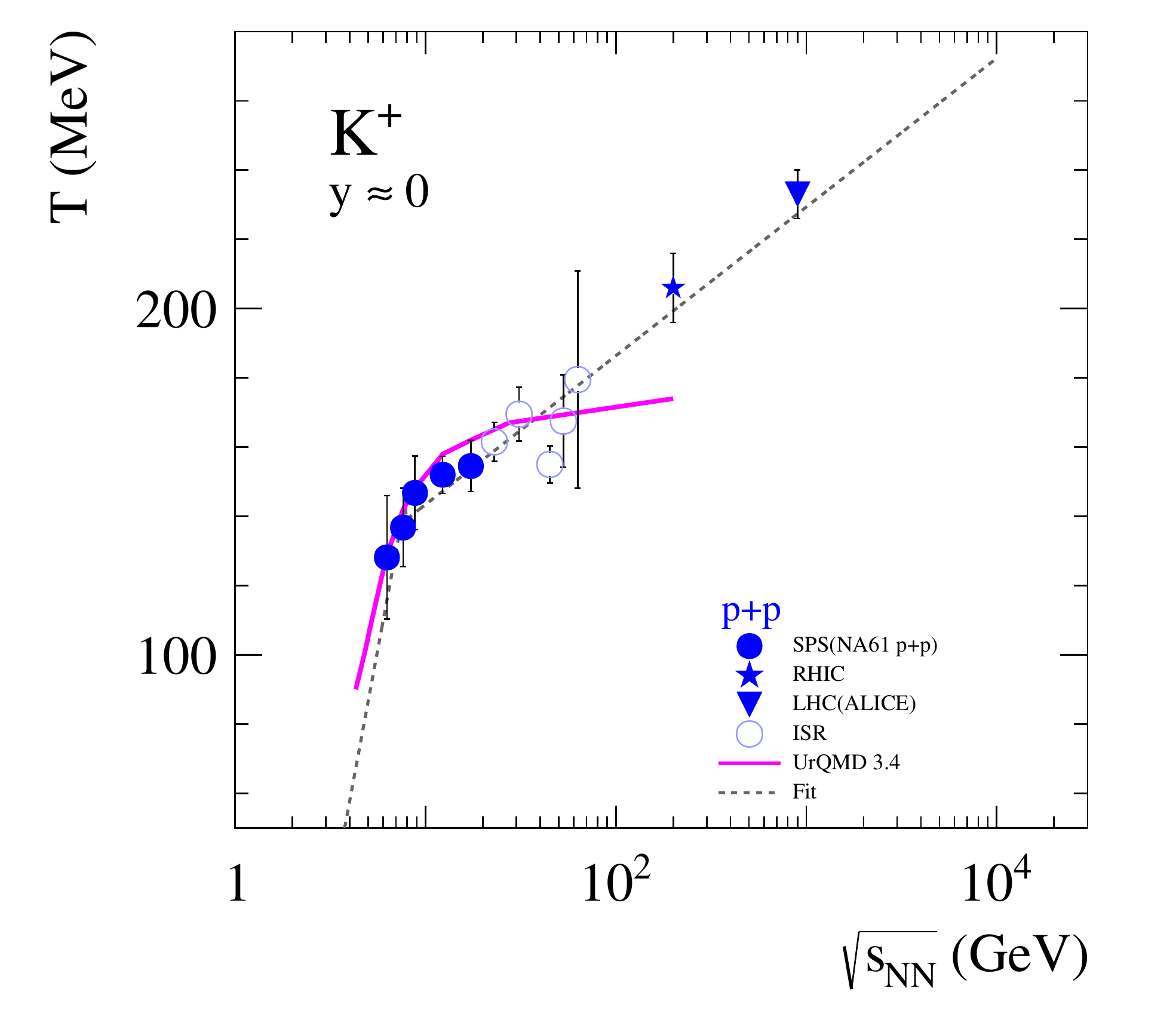}
	\end{center}
	\caption{Energy dependence of the \kp/\pip ratio in inelastic p+p interactions  
		at mid-rapidity (\emph{top-left}) and in the full phase-space (\emph{top-right}) as well 
		the inverse slope parameter $T$ of transverse mass spectra at mid-rapidity for \km (\emph{bottom-left})
		and \kp  (\emph{bottom-right}) mesons. The data are fitted by two straight lines in order to locate a position of the break in the energy dependence. 
	    The experimental results are compared with predictions of the resonance-string model, UrQMD~\protect{\cite{Bleicher:1999xi}}.}
	\label{fig:pp}
%		\end{minipage}
%}
\end{figure*}

To estimate the break energy between a fast rise at low energies and a plateau or slower increase at 
high energies two straight lines were fitted to the p+p data (see Fig.~\ref{fig:pp}). 
The low energy line was constrained by the threshold energy for kaon production.
The fitted break energy is $8.3\pm0.6$~\GeV, $7.70\pm0.14$~\GeV,  $6.5\pm0.5$~\GeV
and $7.9\pm0.2$~\GeV, for the \kp/\pip, $\langle \kp \rangle/\langle \pip \rangle$
ratios and $T(\km)$, $T(\kp)$, respectively. These values are close to each other and 
surprisingly close to 
the energy of the beginning of the horn and step structures in central Pb+Pb collisions - the transition energy 
being approximately 8~\GeV (see Fig.~\ref{fig:ppPbPb}).

%%% UrQMD
Figure~\ref{fig:pp} presents also predictions of a resonance-string model, UrQMD~\cite{Bleicher:1999xi}. 
This model assumes the transition between particle production by resonance formation at low 
energies and string formation at high energies~\cite{Vovchenko:2014vda} - \textit{the resonance-string transition}.
The postulate allows to approximately fit the fast low-energy increase of the \kp/\pip ratio and 
the inverse slope parameter of transverse mass spectra of charged kaons and its slowing down at high energies.
The sharpness of the break is not reproduced by the model.

The unexpected similarity of the transition energy in central Pb+Pb collisions
and the break energy in p+p interactions provokes the question whether there is a common physics origin of the
two effects. 
Is this coincidence accidental? 
If not, do we see effects included in standard modelling of heavy-ion collisions in p+p interactions, or reversely 
non-equilibrium processes in
p+p interactions lead to the horn and step in heavy-ion collisions?
The recent LHC results on hydrodynamical properties of p+p interactions suggest the former interpretation.
However, the discussed experimental results are brand new and the community is far from reaching a consensus.
An obstacle is that the validity of quantitative models is usually restricted to a limited range in collision energy, size of colliding nuclei and concerns only selected observables. Examples of recent developments are given in
Refs.~\cite{Cassing:2015owa,Batyuk:2016qmb,Weil:2016zrk,Jalilian-Marian:2019xgc}.

In summary, new results of \NASixtyOne on the collision energy dependence of the \kp/\pip ratio and the inverse slope
parameter of kaon \mt spectra in inelastic p+p interactions 
are presented together with a compilation of the world data.
The p+p results are compared with the corresponding measurements in central Pb+Pb and Au+Au collisions.
The comparison uncovers a similarity between the collision energy dependence 
in p+p interactions and central heavy ion collisions -
a rapid change of collision energy dependence of basic hadron production properties in the same energy range. 
Possible interpretations are briefly discussed. 
Clearly, understanding of the origin for the similarity between results on heavy ion collisions and p+p interactions
is one of the key objectives of heavy ion physics today.
Emerging results from the \NASixtyOne (nuclear mass number) - (collision energy) scan as well as results from
LHC and RHIC on collisions of small and medium size systems qualitatively change the experimental landscape.
In parallel, significant progress is needed in the modelling of collision energy and system size dependence  
which would extend the validity of models to the full range covered by the data.   

\begin{acknowledgments}
We would like to thank the CERN EP, BE, HSE and EN Departments for the
strong support of NA61/SHINE.

This work was supported by
the Hungarian Scientific Research Fund (grant NKFIH 123842\slash123959),
the Polish Ministry of Science
and Higher Education (grants 667\slash N-CERN\slash2010\slash0,
NN\,202\,48\,4339 and NN\,202\,23\,1837), the National Science Centre Poland (grants~2011\slash03\slash N\slash ST2\slash03691,
2013\slash11\slash N\slash ST2\slash03879, 2014\slash13\slash N\slash
ST2\slash02565, 2014\slash14\slash E\slash ST2\slash00018,
2014\slash15\slash B\slash ST2\slash02537 and
2015\slash18\slash M\slash ST2\slash00125, 2015\slash 19\slash N\slash ST2 \slash01689, 2016\slash23\slash B\slash ST2\slash00692, 
2017\slash 25\slash N\slash ST2\slash 02575, 2018\slash 30\slash A\slash ST2\slash 00226),
the Russian Science Foundation, grant 16-12-10176,
the Russian Academy of Science and the
Russian Foundation for Basic Research (grants 08-02-00018, 09-02-00664
and 12-02-91503-CERN), the Ministry of Science and
Education of the Russian Federation, grant No.\ 3.3380.2017\slash4.6,
 the National Research Nuclear
University MEPhI in the framework of the Russian Academic Excellence
Project (contract No.\ 02.a03.21.0005, 27.08.2013),
the Ministry of Education, Culture, Sports,
Science and Tech\-no\-lo\-gy, Japan, Grant-in-Aid for Sci\-en\-ti\-fic
Research (grants 18071005, 19034011, 19740162, 20740160 and 20039012),
the German Research Foundation (grant GA\,1480/2-2), the
Bulgarian Nuclear Regulatory Agency and the Joint Institute for
Nuclear Research, Dubna (bilateral contract No. 4799-1-18\slash 20),
Bulgarian National Science Fund (grant DN08/11), Ministry of Education
and Science of the Republic of Serbia (grant OI171002), Swiss
Nationalfonds Foundation (grant 200020\-117913/1), ETH Research Grant
TH-01\,07-3 and the Fermi National Accelerator Laboratory (Fermilab), a U.S. Department of Energy, Office of Science, HEP User Facility. Fermilab is managed by Fermi Research Alliance, LLC (FRA), acting under Contract No. DE-AC02-07CH11359.

\end{acknowledgments}

\bibliography{na61References}

\begin{thebibliography}{46}
\expandafter\ifx\csname natexlab\endcsname\relax\def\natexlab#1{#1}\fi
\expandafter\ifx\csname bibnamefont\endcsname\relax
  \def\bibnamefont#1{#1}\fi
\expandafter\ifx\csname bibfnamefont\endcsname\relax
  \def\bibfnamefont#1{#1}\fi
\expandafter\ifx\csname citenamefont\endcsname\relax
  \def\citenamefont#1{#1}\fi
\expandafter\ifx\csname url\endcsname\relax
  \def\url#1{\texttt{#1}}\fi
\expandafter\ifx\csname urlprefix\endcsname\relax\def\urlprefix{URL }\fi
\providecommand{\bibinfo}[2]{#2}
\providecommand{\eprint}[2][]{\url{#2}}

\bibitem[{\citenamefont{Florkowski}(2010)}]{Florkowski:2010zz}
\bibinfo{author}{\bibfnamefont{W.}~\bibnamefont{Florkowski}},
  \emph{\bibinfo{title}{{Phenomenology of Ultra-Relativistic Heavy-Ion
  Collisions}}} (\bibinfo{year}{2010}), ISBN \bibinfo{isbn}{9789814280662}.

\bibitem[{\citenamefont{Torrieri et~al.}(2005)\citenamefont{Torrieri, Steinke,
  Broniowski, Florkowski, Letessier, and Rafelski}}]{Torrieri:2004zz}
\bibinfo{author}{\bibfnamefont{G.}~\bibnamefont{Torrieri}},
  \bibinfo{author}{\bibfnamefont{S.}~\bibnamefont{Steinke}},
  \bibinfo{author}{\bibfnamefont{W.}~\bibnamefont{Broniowski}},
  \bibinfo{author}{\bibfnamefont{W.}~\bibnamefont{Florkowski}},
  \bibinfo{author}{\bibfnamefont{J.}~\bibnamefont{Letessier}},
  \bibnamefont{and} \bibinfo{author}{\bibfnamefont{J.}~\bibnamefont{Rafelski}},
  \bibinfo{journal}{Comput. Phys. Commun.} \textbf{\bibinfo{volume}{167}},
  \bibinfo{pages}{229} (\bibinfo{year}{2005}), \eprint{nucl-th/0404083}.

\bibitem[{\citenamefont{Becattini et~al.}(2006)\citenamefont{Becattini,
  Manninen, and Gazdzicki}}]{Becattini:2005xt}
\bibinfo{author}{\bibfnamefont{F.}~\bibnamefont{Becattini}},
  \bibinfo{author}{\bibfnamefont{J.}~\bibnamefont{Manninen}}, \bibnamefont{and}
  \bibinfo{author}{\bibfnamefont{M.}~\bibnamefont{Gazdzicki}},
  \bibinfo{journal}{Phys.Rev.} \textbf{\bibinfo{volume}{C73}},
  \bibinfo{pages}{044905} (\bibinfo{year}{2006}), \eprint{hep-ph/0511092}.

\bibitem[{\citenamefont{Andronic et~al.}(2018)\citenamefont{Andronic,
  Braun-Munzinger, Redlich, and Stachel}}]{Andronic:2017pug}
\bibinfo{author}{\bibfnamefont{A.}~\bibnamefont{Andronic}},
  \bibinfo{author}{\bibfnamefont{P.}~\bibnamefont{Braun-Munzinger}},
  \bibinfo{author}{\bibfnamefont{K.}~\bibnamefont{Redlich}}, \bibnamefont{and}
  \bibinfo{author}{\bibfnamefont{J.}~\bibnamefont{Stachel}},
  \bibinfo{journal}{Nature} \textbf{\bibinfo{volume}{561}},
  \bibinfo{pages}{321} (\bibinfo{year}{2018}), \eprint{1710.09425}.

\bibitem[{\citenamefont{Afanasiev et~al.}(2002)}]{Afanasiev:2002mx}
\bibinfo{author}{\bibfnamefont{S.}~\bibnamefont{Afanasiev}}
  \bibnamefont{et~al.} (\bibinfo{collaboration}{NA49}),
  \bibinfo{journal}{Phys.\ Rev.} \textbf{\bibinfo{volume}{C66}},
  \bibinfo{pages}{054902} (\bibinfo{year}{2002}).

\bibitem[{\citenamefont{Alt et~al.}(2008)}]{Alt:2007aa}
\bibinfo{author}{\bibfnamefont{C.}~\bibnamefont{Alt}} \bibnamefont{et~al.}
  (\bibinfo{collaboration}{NA49}), \bibinfo{journal}{Phys.\ Rev.}
  \textbf{\bibinfo{volume}{C77}}, \bibinfo{pages}{024903}
  (\bibinfo{year}{2008}).

\bibitem[{\citenamefont{Gazdzicki et~al.}(2011)\citenamefont{Gazdzicki,
  Gorenstein, and Seyboth}}]{Gazdzicki:2010iv}
\bibinfo{author}{\bibfnamefont{M.}~\bibnamefont{Gazdzicki}},
  \bibinfo{author}{\bibfnamefont{M.}~\bibnamefont{Gorenstein}},
  \bibnamefont{and} \bibinfo{author}{\bibfnamefont{P.}~\bibnamefont{Seyboth}},
  \bibinfo{journal}{Acta Phys.Polon.} \textbf{\bibinfo{volume}{B42}},
  \bibinfo{pages}{307} (\bibinfo{year}{2011}), \eprint{1006.1765}.

\bibitem[{\citenamefont{Gazdzicki and Gorenstein}(1999)}]{Gazdzicki:1998vd}
\bibinfo{author}{\bibfnamefont{M.}~\bibnamefont{Gazdzicki}} \bibnamefont{and}
  \bibinfo{author}{\bibfnamefont{M.~I.} \bibnamefont{Gorenstein}},
  \bibinfo{journal}{Acta Phys.Polon.} \textbf{\bibinfo{volume}{B30}},
  \bibinfo{pages}{2705} (\bibinfo{year}{1999}), \eprint{hep-ph/9803462}.

\bibitem[{\citenamefont{Andersson et~al.}(1983)\citenamefont{Andersson,
  Gustafson, Ingelman, and Sjostrand}}]{Andersson:1983ia}
\bibinfo{author}{\bibfnamefont{B.}~\bibnamefont{Andersson}},
  \bibinfo{author}{\bibfnamefont{G.}~\bibnamefont{Gustafson}},
  \bibinfo{author}{\bibfnamefont{G.}~\bibnamefont{Ingelman}}, \bibnamefont{and}
  \bibinfo{author}{\bibfnamefont{T.}~\bibnamefont{Sjostrand}},
  \bibinfo{journal}{Phys. Rept.} \textbf{\bibinfo{volume}{97}},
  \bibinfo{pages}{31} (\bibinfo{year}{1983}).

\bibitem[{\citenamefont{Begun et~al.}(2008)\citenamefont{Begun, Gazdzicki, and
  Gorenstein}}]{Begun:2008fm}
\bibinfo{author}{\bibfnamefont{V.~V.} \bibnamefont{Begun}},
  \bibinfo{author}{\bibfnamefont{M.}~\bibnamefont{Gazdzicki}},
  \bibnamefont{and} \bibinfo{author}{\bibfnamefont{M.~I.}
  \bibnamefont{Gorenstein}}, \bibinfo{journal}{Phys. Rev.}
  \textbf{\bibinfo{volume}{C78}}, \bibinfo{pages}{024904}
  (\bibinfo{year}{2008}), \eprint{0804.0075}.

\bibitem[{\citenamefont{Chatrchyan et~al.}(2013)}]{CMS:2012qk}
\bibinfo{author}{\bibfnamefont{S.}~\bibnamefont{Chatrchyan}}
  \bibnamefont{et~al.} (\bibinfo{collaboration}{CMS}),
  \bibinfo{journal}{Phys.Lett.} \textbf{\bibinfo{volume}{B718}},
  \bibinfo{pages}{795} (\bibinfo{year}{2013}), \eprint{1210.5482}.

\bibitem[{\citenamefont{Khachatryan et~al.}(2017)}]{Khachatryan:2016txc}
\bibinfo{author}{\bibfnamefont{V.}~\bibnamefont{Khachatryan}}
  \bibnamefont{et~al.} (\bibinfo{collaboration}{CMS}), \bibinfo{journal}{Phys.
  Lett.} \textbf{\bibinfo{volume}{B765}}, \bibinfo{pages}{193}
  (\bibinfo{year}{2017}), \eprint{1606.06198}.

\bibitem[{\citenamefont{Nagle and Zajc}(2018)}]{Nagle:2018nvi}
\bibinfo{author}{\bibfnamefont{J.~L.} \bibnamefont{Nagle}} \bibnamefont{and}
  \bibinfo{author}{\bibfnamefont{W.~A.} \bibnamefont{Zajc}},
  \bibinfo{journal}{Ann. Rev. Nucl. Part. Sci.} \textbf{\bibinfo{volume}{68}},
  \bibinfo{pages}{211} (\bibinfo{year}{2018}), \eprint{1801.03477}.

\bibitem[{\citenamefont{Bozek and Broniowski}(2016)}]{Bozek:2016acp}
\bibinfo{author}{\bibfnamefont{P.}~\bibnamefont{Bozek}} \bibnamefont{and}
  \bibinfo{author}{\bibfnamefont{W.}~\bibnamefont{Broniowski}},
  \bibinfo{journal}{PoS} \textbf{\bibinfo{volume}{LHCP2016}},
  \bibinfo{pages}{116} (\bibinfo{year}{2016}).

\bibitem[{\citenamefont{Adam et~al.}(2017)}]{ALICE:2017jyt}
\bibinfo{author}{\bibfnamefont{J.}~\bibnamefont{Adam}} \bibnamefont{et~al.}
  (\bibinfo{collaboration}{ALICE}), \bibinfo{journal}{Nature Phys.}
  \textbf{\bibinfo{volume}{13}}, \bibinfo{pages}{535} (\bibinfo{year}{2017}),
  \eprint{1606.07424}.

\bibitem[{\citenamefont{Abgrall et~al.}(2014)}]{Abgrall:2014fa}
\bibinfo{author}{\bibfnamefont{N.}~\bibnamefont{Abgrall}} \bibnamefont{et~al.}
  (\bibinfo{collaboration}{NA61/SHINE}), \bibinfo{journal}{JINST}
  \textbf{\bibinfo{volume}{9}}, \bibinfo{pages}{P06005} (\bibinfo{year}{2014}),
  \eprint{1401.4699}.

\bibitem[{\citenamefont{Aduszkiewicz et~al.}(2017)}]{Aduszkiewicz:2017sei}
\bibinfo{author}{\bibfnamefont{A.}~\bibnamefont{Aduszkiewicz}}
  \bibnamefont{et~al.} (\bibinfo{collaboration}{NA61/SHINE}),
  \bibinfo{journal}{European Physical Journal} \textbf{\bibinfo{volume}{C77}},
  \bibinfo{pages}{671} (\bibinfo{year}{2017}), \eprint{1705.02467}.

\bibitem[{\citenamefont{Aduszkiewicz et~al.}(2016)}]{Aduszkiewicz:2015jna}
\bibinfo{author}{\bibfnamefont{A.}~\bibnamefont{Aduszkiewicz}}
  \bibnamefont{et~al.} (\bibinfo{collaboration}{NA61/SHINE}),
  \bibinfo{journal}{Eur. Phys. J.} \textbf{\bibinfo{volume}{C76}},
  \bibinfo{pages}{635} (\bibinfo{year}{2016}), \eprint{1510.00163}.

\bibitem[{\citenamefont{Gazdzicki and Roehrich}(1995)}]{Gazdzicki:1995zs}
\bibinfo{author}{\bibfnamefont{M.}~\bibnamefont{Gazdzicki}} \bibnamefont{and}
  \bibinfo{author}{\bibfnamefont{D.}~\bibnamefont{Roehrich}},
  \bibinfo{journal}{Z.Phys.} \textbf{\bibinfo{volume}{C65}},
  \bibinfo{pages}{215} (\bibinfo{year}{1995}).

\bibitem[{\citenamefont{Gazdzicki and Rohrich}(1996)}]{Gazdzicki:1996pk}
\bibinfo{author}{\bibfnamefont{M.}~\bibnamefont{Gazdzicki}} \bibnamefont{and}
  \bibinfo{author}{\bibfnamefont{D.}~\bibnamefont{Rohrich}},
  \bibinfo{journal}{Z.Phys.} \textbf{\bibinfo{volume}{C71}},
  \bibinfo{pages}{55} (\bibinfo{year}{1996}), \eprint{hep-ex/9607004}.

\bibitem[{\citenamefont{Kliemant et~al.}(2004)\citenamefont{Kliemant, Lungwitz,
  and Gazdzicki}}]{Kliemant:2003sa}
\bibinfo{author}{\bibfnamefont{M.}~\bibnamefont{Kliemant}},
  \bibinfo{author}{\bibfnamefont{B.}~\bibnamefont{Lungwitz}}, \bibnamefont{and}
  \bibinfo{author}{\bibfnamefont{M.}~\bibnamefont{Gazdzicki}},
  \bibinfo{journal}{Phys.Rev.} \textbf{\bibinfo{volume}{C69}},
  \bibinfo{pages}{044903} (\bibinfo{year}{2004}), \eprint{hep-ex/0308002}.

\bibitem[{\citenamefont{Arsene et~al.}(2005)}]{Arsene:2005mr}
\bibinfo{author}{\bibfnamefont{I.}~\bibnamefont{Arsene}} \bibnamefont{et~al.}
  (\bibinfo{collaboration}{BRAHMS Collaboration}), \bibinfo{journal}{Phys.Rev.}
  \textbf{\bibinfo{volume}{C72}}, \bibinfo{pages}{014908}
  (\bibinfo{year}{2005}), \eprint{nucl-ex/0503010}.

\bibitem[{\citenamefont{Aamodt et~al.}(2011)}]{Aamodt:2011zj}
\bibinfo{author}{\bibfnamefont{K.}~\bibnamefont{Aamodt}} \bibnamefont{et~al.}
  (\bibinfo{collaboration}{ALICE Collaboration}),
  \bibinfo{journal}{Eur.Phys.J.} \textbf{\bibinfo{volume}{C71}},
  \bibinfo{pages}{1655} (\bibinfo{year}{2011}), \eprint{1101.4110}.

\bibitem[{\citenamefont{Abelev et~al.}(2012)}]{Abelev:2012wca}
\bibinfo{author}{\bibfnamefont{B.}~\bibnamefont{Abelev}} \bibnamefont{et~al.}
  (\bibinfo{collaboration}{ALICE Collaboration}),
  \bibinfo{journal}{Phys.Rev.Lett.} \textbf{\bibinfo{volume}{109}},
  \bibinfo{pages}{252301} (\bibinfo{year}{2012}), \eprint{1208.1974}.

\bibitem[{\citenamefont{Alper et~al.}(1975)}]{Alper:1975jm}
\bibinfo{author}{\bibfnamefont{B.}~\bibnamefont{Alper}} \bibnamefont{et~al.}
  (\bibinfo{collaboration}{British-Scandinavian}), \bibinfo{journal}{Nucl.
  Phys.} \textbf{\bibinfo{volume}{B100}}, \bibinfo{pages}{237}
  (\bibinfo{year}{1975}).

\bibitem[{\citenamefont{Abelev et~al.}(2009)}]{Abelev:2008ab}
\bibinfo{author}{\bibfnamefont{B.}~\bibnamefont{Abelev}} \bibnamefont{et~al.}
  (\bibinfo{collaboration}{STAR Collaboration}), \bibinfo{journal}{Phys.Rev.}
  \textbf{\bibinfo{volume}{C79}}, \bibinfo{pages}{034909}
  (\bibinfo{year}{2009}), \eprint{0808.2041}.

\bibitem[{\citenamefont{Abelev et~al.}(2014)}]{Abelev:2014laa}
\bibinfo{author}{\bibfnamefont{B.~B.} \bibnamefont{Abelev}}
  \bibnamefont{et~al.} (\bibinfo{collaboration}{ALICE Collaboration}),
  \bibinfo{journal}{Phys.Lett.} \textbf{\bibinfo{volume}{B736}},
  \bibinfo{pages}{196} (\bibinfo{year}{2014}), \eprint{1401.1250}.

\bibitem[{\citenamefont{Ahle et~al.}(1998{\natexlab{a}})}]{Ahle:1998jc}
\bibinfo{author}{\bibfnamefont{L.}~\bibnamefont{Ahle}} \bibnamefont{et~al.}
  (\bibinfo{collaboration}{E802}), \bibinfo{journal}{Phys.\ Rev.}
  \textbf{\bibinfo{volume}{C57}}, \bibinfo{pages}{466}
  (\bibinfo{year}{1998}{\natexlab{a}}).

\bibitem[{\citenamefont{Ahle et~al.}(1998{\natexlab{b}})}]{Ahle:1998gv}
\bibinfo{author}{\bibfnamefont{L.}~\bibnamefont{Ahle}} \bibnamefont{et~al.}
  (\bibinfo{collaboration}{E-802}), \bibinfo{journal}{Phys. Rev.}
  \textbf{\bibinfo{volume}{C58}}, \bibinfo{pages}{3523}
  (\bibinfo{year}{1998}{\natexlab{b}}).

\bibitem[{\citenamefont{Ahle et~al.}(1999)}]{Ahle:1999va}
\bibinfo{author}{\bibfnamefont{L.}~\bibnamefont{Ahle}} \bibnamefont{et~al.}
  (\bibinfo{collaboration}{E-802, E-866}), \bibinfo{journal}{Phys. Rev.}
  \textbf{\bibinfo{volume}{C60}}, \bibinfo{pages}{044904}
  (\bibinfo{year}{1999}), \eprint{nucl-ex/9903009}.

\bibitem[{\citenamefont{Ahle et~al.}(2000{\natexlab{a}})}]{Ahle:1999uy}
\bibinfo{author}{\bibfnamefont{L.}~\bibnamefont{Ahle}} \bibnamefont{et~al.}
  (\bibinfo{collaboration}{E866, E917}), \bibinfo{journal}{Phys. Lett.}
  \textbf{\bibinfo{volume}{B476}}, \bibinfo{pages}{1}
  (\bibinfo{year}{2000}{\natexlab{a}}), \eprint{nucl-ex/9910008}.

\bibitem[{\citenamefont{Ahle et~al.}(2000{\natexlab{b}})}]{Ahle:2000wq}
\bibinfo{author}{\bibfnamefont{L.}~\bibnamefont{Ahle}} \bibnamefont{et~al.}
  (\bibinfo{collaboration}{E866, E917}), \bibinfo{journal}{Phys. Lett.}
  \textbf{\bibinfo{volume}{B490}}, \bibinfo{pages}{53}
  (\bibinfo{year}{2000}{\natexlab{b}}), \eprint{nucl-ex/0008010}.

\bibitem[{\citenamefont{Barrette et~al.}(2000)}]{Barrette:1999ry}
\bibinfo{author}{\bibfnamefont{J.}~\bibnamefont{Barrette}} \bibnamefont{et~al.}
  (\bibinfo{collaboration}{E877}), \bibinfo{journal}{Phys. Rev.}
  \textbf{\bibinfo{volume}{C62}}, \bibinfo{pages}{024901}
  (\bibinfo{year}{2000}), \eprint{nucl-ex/9910004}.

\bibitem[{\citenamefont{Pelte et~al.}(1997)}]{Pelte:1997rg}
\bibinfo{author}{\bibfnamefont{D.}~\bibnamefont{Pelte}} \bibnamefont{et~al.}
  (\bibinfo{collaboration}{FOPI}), \bibinfo{journal}{Z. Phys.}
  \textbf{\bibinfo{volume}{A357}}, \bibinfo{pages}{215} (\bibinfo{year}{1997}).

\bibitem[{\citenamefont{Bearden et~al.}(2005)}]{Bearden:2004yx}
\bibinfo{author}{\bibfnamefont{I.}~\bibnamefont{Bearden}} \bibnamefont{et~al.}
  (\bibinfo{collaboration}{BRAHMS}), \bibinfo{journal}{Phys.\ Rev.\ Lett.}
  \textbf{\bibinfo{volume}{94}}, \bibinfo{pages}{162301}
  (\bibinfo{year}{2005}).

\bibitem[{\citenamefont{Adamczyk et~al.}(2017)}]{Adamczyk:2017iwn}
\bibinfo{author}{\bibfnamefont{L.}~\bibnamefont{Adamczyk}} \bibnamefont{et~al.}
  (\bibinfo{collaboration}{STAR}), \bibinfo{journal}{Phys. Rev.}
  \textbf{\bibinfo{volume}{C96}}, \bibinfo{pages}{044904}
  (\bibinfo{year}{2017}), \eprint{1701.07065}.

\bibitem[{\citenamefont{Abelev et~al.}(2013)}]{Abelev:2013vea}
\bibinfo{author}{\bibfnamefont{B.}~\bibnamefont{Abelev}} \bibnamefont{et~al.}
  (\bibinfo{collaboration}{ALICE}), \bibinfo{journal}{Phys. Rev.}
  \textbf{\bibinfo{volume}{C88}}, \bibinfo{pages}{044910}
  (\bibinfo{year}{2013}), \eprint{1303.0737}.

\bibitem[{\citenamefont{Adam et~al.}(2015)}]{Adam:2015qaa}
\bibinfo{author}{\bibfnamefont{J.}~\bibnamefont{Adam}} \bibnamefont{et~al.}
  (\bibinfo{collaboration}{ALICE}), \bibinfo{journal}{Eur. Phys. J.}
  \textbf{\bibinfo{volume}{C75}}, \bibinfo{pages}{226} (\bibinfo{year}{2015}),
  \eprint{1504.00024}.

\bibitem[{\citenamefont{Koch et~al.}(1986)\citenamefont{Koch, Muller, and
  Rafelski}}]{Koch:1986ud}
\bibinfo{author}{\bibfnamefont{P.}~\bibnamefont{Koch}},
  \bibinfo{author}{\bibfnamefont{B.}~\bibnamefont{Muller}}, \bibnamefont{and}
  \bibinfo{author}{\bibfnamefont{J.}~\bibnamefont{Rafelski}},
  \bibinfo{journal}{Phys. Rept.} \textbf{\bibinfo{volume}{142}},
  \bibinfo{pages}{167} (\bibinfo{year}{1986}).

\bibitem[{\citenamefont{Gorenstein et~al.}(2003)\citenamefont{Gorenstein,
  Gazdzicki, and Bugaev}}]{Gorenstein:2003cu}
\bibinfo{author}{\bibfnamefont{M.~I.} \bibnamefont{Gorenstein}},
  \bibinfo{author}{\bibfnamefont{M.}~\bibnamefont{Gazdzicki}},
  \bibnamefont{and} \bibinfo{author}{\bibfnamefont{K.~A.}
  \bibnamefont{Bugaev}}, \bibinfo{journal}{Phys. Lett.}
  \textbf{\bibinfo{volume}{B567}}, \bibinfo{pages}{175} (\bibinfo{year}{2003}),
  \eprint{hep-ph/0303041}.

\bibitem[{\citenamefont{Bleicher et~al.}(1999)}]{Bleicher:1999xi}
\bibinfo{author}{\bibfnamefont{M.}~\bibnamefont{Bleicher}}
  \bibnamefont{et~al.}, \bibinfo{journal}{J.Phys.}
  \textbf{\bibinfo{volume}{G25}}, \bibinfo{pages}{1859} (\bibinfo{year}{1999}),
  \eprint{hep-ph/9909407}.

\bibitem[{\citenamefont{Vovchenko et~al.}(2015)\citenamefont{Vovchenko,
  Anchishkin, and Gorenstein}}]{Vovchenko:2014vda}
\bibinfo{author}{\bibfnamefont{V.~{\relax Yu}.} \bibnamefont{Vovchenko}},
  \bibinfo{author}{\bibfnamefont{D.~V.} \bibnamefont{Anchishkin}},
  \bibnamefont{and} \bibinfo{author}{\bibfnamefont{M.~I.}
  \bibnamefont{Gorenstein}}, \bibinfo{journal}{Nucl. Phys.}
  \textbf{\bibinfo{volume}{A936}}, \bibinfo{pages}{1} (\bibinfo{year}{2015}),
  \eprint{1408.5493}.

\bibitem[{\citenamefont{Cassing et~al.}(2016)\citenamefont{Cassing, Palmese,
  Moreau, and Bratkovskaya}}]{Cassing:2015owa}
\bibinfo{author}{\bibfnamefont{W.}~\bibnamefont{Cassing}},
  \bibinfo{author}{\bibfnamefont{A.}~\bibnamefont{Palmese}},
  \bibinfo{author}{\bibfnamefont{P.}~\bibnamefont{Moreau}}, \bibnamefont{and}
  \bibinfo{author}{\bibfnamefont{E.~L.} \bibnamefont{Bratkovskaya}},
  \bibinfo{journal}{Phys. Rev.} \textbf{\bibinfo{volume}{C93}},
  \bibinfo{pages}{014902} (\bibinfo{year}{2016}), \eprint{1510.04120}.

\bibitem[{\citenamefont{Batyuk et~al.}(2016)\citenamefont{Batyuk, Blaschke,
  Bleicher, Ivanov, Karpenko, Merts, Nahrgang, Petersen, and
  Rogachevsky}}]{Batyuk:2016qmb}
\bibinfo{author}{\bibfnamefont{P.}~\bibnamefont{Batyuk}},
  \bibinfo{author}{\bibfnamefont{D.}~\bibnamefont{Blaschke}},
  \bibinfo{author}{\bibfnamefont{M.}~\bibnamefont{Bleicher}},
  \bibinfo{author}{\bibfnamefont{{\relax Yu}.~B.} \bibnamefont{Ivanov}},
  \bibinfo{author}{\bibfnamefont{I.}~\bibnamefont{Karpenko}},
  \bibinfo{author}{\bibfnamefont{S.}~\bibnamefont{Merts}},
  \bibinfo{author}{\bibfnamefont{M.}~\bibnamefont{Nahrgang}},
  \bibinfo{author}{\bibfnamefont{H.}~\bibnamefont{Petersen}}, \bibnamefont{and}
  \bibinfo{author}{\bibfnamefont{O.}~\bibnamefont{Rogachevsky}},
  \bibinfo{journal}{Phys. Rev.} \textbf{\bibinfo{volume}{C94}},
  \bibinfo{pages}{044917} (\bibinfo{year}{2016}), \eprint{1608.00965}.

\bibitem[{\citenamefont{Weil et~al.}(2016)}]{Weil:2016zrk}
\bibinfo{author}{\bibfnamefont{J.}~\bibnamefont{Weil}} \bibnamefont{et~al.},
  \bibinfo{journal}{Phys. Rev.} \textbf{\bibinfo{volume}{C94}},
  \bibinfo{pages}{054905} (\bibinfo{year}{2016}), \eprint{1606.06642}.

\bibitem[{\citenamefont{Jalilian-Marian}(2019)}]{Jalilian-Marian:2019xgc}
\bibinfo{author}{\bibfnamefont{J.}~\bibnamefont{Jalilian-Marian}},
  \bibinfo{journal}{Nucl. Phys.} \textbf{\bibinfo{volume}{A982}},
  \bibinfo{pages}{935} (\bibinfo{year}{2019}).

\end{thebibliography}
\end{document}